\documentclass[conf]{new-aiaa}
\usepackage[utf8]{inputenc}
\usepackage[T1]{fontenc}
\usepackage{graphicx}
\usepackage{amsmath}
\usepackage[version=4]{mhchem}
\usepackage{siunitx}
\usepackage{longtable,tabularx}
\usepackage{ragged2e}
\usepackage{colortbl}
\usepackage{float}
\usepackage{multicol}
\usepackage{algorithm} 
\usepackage{algpseudocode}

\title{Estimating Probability Density Function \\of Vehicle-Related Time Series Data using \\Histogram Cubic B-Spline Approximation}

\author{Nico Schick\footnote{N. Schick, M. Sc. studied Applied Computer Sciences (M. Sc.) and Computer Engineering (B. Eng.) at the Esslingen University of Applied Sciences. e-mail: Nico.Schick@hs-esslingen.de}}

\begin{document}

\maketitle

\begin{abstract}
Random processes play a crucial role in scientific research, often characterized by distribution functions or probability density functions (PDFs). These PDFs serve as essential approximations of the actual and frequently undisclosed distribution governing the random process under examination. Diverse methodologies exist for estimating PDFs, each offering distinct advantages in specific contexts. This publication presents a novel approach that centers on estimating probability density functions by leveraging histograms and B-spline curves, with a particular focus on analyzing vehicle-related time series data. The proposed method outlines a comprehensive framework for estimating PDFs tailored specifically to the study of vehicle-related phenomena. By effectively combining the strengths of histograms and B-spline curves, researchers gain a powerful toolset to obtain precise and reliable estimations of PDFs, thereby enabling advanced analysis and comprehension of vehicle-related random processes in scientific investigations.
\end{abstract}

\textbf{Scientific Question:} \textit{How probability density functions can be estimated for vehicle-related time series data based on histograms and B-spline curves?}


\section{Motivation}
Random processes are of paramount importance in scientific research, as they often underpin the behavior of various phenomena. Understanding these processes requires the characterization of their distribution functions or probability density functions (PDFs). PDFs serve as crucial approximations of the true, but frequently unknown, distributions governing the random processes under investigation. Accurate estimation of PDFs is therefore vital for gaining insights into the underlying dynamics and making informed decisions based on the observed data. \cite{Mot1}

The estimation of PDFs plays a fundamental role in numerous scientific disciplines. In fields such as statistics, engineering, finance, and environmental science, PDFs are used to model and analyze complex systems and phenomena. By capturing the statistical properties of the observed data, PDFs provide a concise summary of the probability distribution and enable various statistical analyses and predictions. Thus, accurate estimation of PDFs is crucial for advancing scientific understanding and facilitating decision-making processes.  \cite{Mot2}

\subsection{Existing Approaches for PDF Estimation}
\label{existsMethods}
Numerous methodologies have been developed for estimating PDFs. For example, based on a naïve estimator \cite{SPDF1}, Bayesian inference \cite{SPDF2}, \textit{Maximum Likelihood Estimation (MLE)} \cite{SPDF3}, \textit{Maximum A Posteriori Probability estimation (MAP)} \cite{SPDF3}, \textit{Method of Moments (MOM)} \cite{SPDF4}, \textit{Empirical Cumulative Distribution Function (ECDF)} \cite{SPDF5}, local polynomial estimation \cite{SPDF6}, \text{k-nearest neighbor method (k-NN)} \cite{SPDF7}, Q-Q plot \cite{SPDF8}, elliptic basis function functions \cite{SPDF9}, neural networks \cite{SPDF10} \cite{SPDF11} \cite{SPDF12}, \textit{Gaussian Mixture Models (GMM)} \cite{SPDF13}, convex optimization \cite{SPDF14}, \textit{Generalized Likelihood Uncertainty Estimation (GLUE)} \cite {SPDF15}, combination of maximum entropy moments and Bayesian method \cite{SPDF16}. Furthermore, probability density functions can be estimated using Kernel Density Estimation (KDE) \cite{SPDF17} \cite{SPDF18} \cite{SPDF19}. Also worth mentioning is a method for estimating the probability density function for relatively small data sets in \cite{SPDF20}. However, these methods may face challenges in capturing the intricate patterns and characteristics of vehicle-related time series datasets.

\subsection{Limitations and the Need for a Novel Approach}
While existing approaches offer valuable insights, there is a need for a novel and tailored method for estimating PDFs in the context of vehicle-related time series data. Vehicle-related phenomena, such as traffic flow patterns, driving behavior, and vehicle performance, exhibit unique characteristics that require specialized analysis techniques. A comprehensive framework is needed to accurately estimate PDFs and unlock the full potential of vehicle-related time series data in scientific investigations. \cite{Mot3}

\subsection{The Proposed Approach: Histogram Cubic B-Spline Approximation}
This publication introduces a novel approach that addresses the limitations of existing methods and focuses specifically on estimating probability density functions of vehicle-related time series data. The proposed method leverages the combined strengths of histograms and B-spline curves to provide precise and reliable estimations of PDFs. By incorporating histogram binning and B-spline approximation techniques, researchers can effectively capture the underlying distribution patterns of vehicle-related processes and gain deeper insights into their dynamics. \cite{Mot4}

\subsection{Advantages and Applications}
The proposed approach offers several advantages over traditional methods. It enables accurate estimation of PDFs by efficiently capturing the complex nature of vehicle-related time series data. Furthermore, it provides a flexible and adaptable framework that can be applied to a wide range of vehicle-related phenomena, including traffic analysis, driver behavior modeling, and performance evaluation. The precise estimation of PDFs using the histogram cubic B-spline approximation empowers researchers to conduct advanced analysis, make informed decisions, and contribute to the advancement of knowledge in the field of vehicle-related studies. \cite{Mot5} 

\newpage
\section{Cubic B-Spline Curve}
\label{sect:spl}
\begin{figure}[H]
\centering
    \includegraphics[scale=0.7]{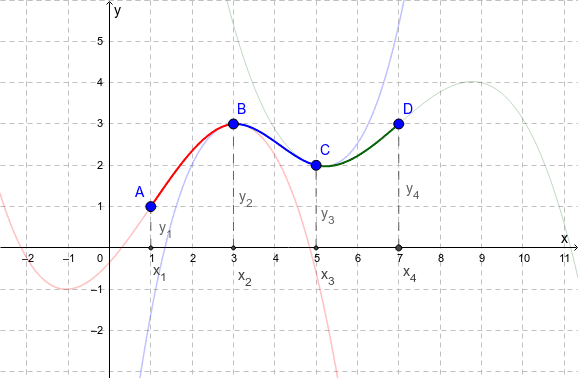}
    \caption{Visualization: Spline curve}
    \label{PubLit}
\end{figure}

A spline is a mathematical function that is made up of several polynomial segments. In the case of an $n$-degree spline, each segment is composed of a polynomial of degree $n$ or lower. The points where two polynomial segments of the spline meet are referred to as nodes. These nodes are arranged or distributed to ensure that the spline is continuously differentiable up to degree $n-1$. Splines are commonly used for interpolating and approximating data points. Their piecewise nature makes them more flexible than polynomials. As a result, splines tend to exhibit less oscillation between the given data points when compared to higher-degree polynomials. Figure \ref{PubLit} provides a visual representation of such a spline curve.

A base spline curve, specifically a cubic base spline curve, is a specialized form of the general spline curve. B-splines, or basis splines, are utilized in the construction of a base spline curve. These B-splines use specific base values or basis functions that are defined over smaller intervals or segments. The combination and interconnection of these B-splines result in the formation of the base spline curve. The advantage of using B-splines is that they offer numerical efficiency and stability in calculations. B-spline curves are mathematical structures that consist of multiple B-splines and rely on a set of essential points known as De Boor points to simplify their definition. In mathematical terms, a cubic B-spline curve can be precisely defined as follows
\begin{align}
	f(u) = \sum_{i=0}^{n-p-2} P_i \, N_{i,p,\tau}(u), \, u \in \left[\tau_p,\tau_{n-p-1}\right]
\end{align}
depending on maximum degree $p$, control points (De Boor Points) $P_i \, (i=0,\hdots,n-p-2)$ and node vector $\tau=(\tau_0,\hdots,\tau_{n-1}) \, \forall \, n \geq 2p$. 

In order to achieve a more numerically efficient calculation of B-splines, they can also be defined recursively as follows:
\begin{align}
	N_{i,0,\tau}(u) = \left\{
	\begin{array}{ll}
		1 & u \in \left[\tau_i,\tau_{i+1}\right] \\
		0 & \, \textrm{else} \\
	\end{array}
	\right.
\end{align}
\label{reqF}
\begin{align}
	N_{i,p,\tau}(u) = \frac{u-\tau_i}{\tau_{i+p}-\tau_i}N_{i,p-1,\tau}(u) + \frac{\tau_{i+p+1}-u}{\tau_{i+p+1}-\tau_{i+1}}N_{i+1,p-1,\tau}(u), \; p>0
\end{align}
The B-spline curve becomes cubic when the parameter $p$ is set to 3. In this case, the curve is influenced solely by the control points $P_i$ within the interval $\left[\tau_i,\tau_{i+p+1}\right]$. To calculate the derivative of a B-spline, the recursion formula mentioned can be applied as follows:
\begin{align}
	\frac{d}{d u} N_{i,p,\tau}(u) = \frac{p}{\tau_{i+p}-\tau_{i}} N_{i,p-1,\tau}(u)-\frac{p}{\tau_{i+p+1}-\tau_{i+1}} N_{i+1,p-1,\tau}(u), \; p \geq 1
\end{align}
A thorough and comprehensive explanation of all the mentioned facts regarding splines and cubic B-splines can be found in \cite{S1}, \cite{S2}, and \cite{S3}. These references provide detailed insights and information on the subject matter.

\section{Histogram Cubic B-Spline Approximation}
\label{subsect:HistoBSplApprox}
In practical situations, it is common for the probability density functions of real processes to be unknown. To tackle this challenge, various approaches and methods are mentioned in Chapter \ref{existsMethods} for estimating probability density functions. An alternative is the use of a histogram cubic B-spline approximation. The fundamentals of this approach itself can be found in \cite{MS}.

\begin{figure}[H]
\centering
    \includegraphics[scale=0.46]{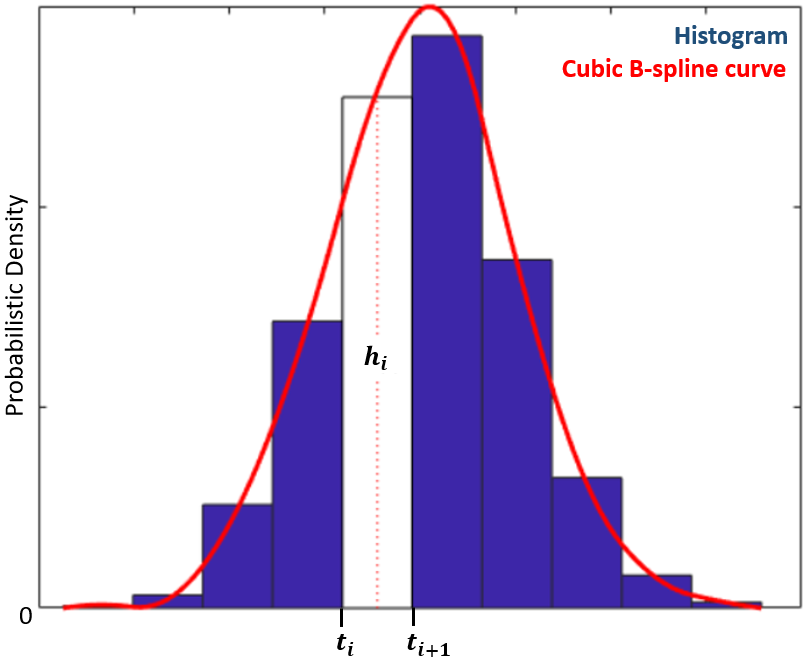}
\caption{Histogram Cubic B-Spline Approximation}
\label{fig:Histo}
\end{figure}

The histogram cubic B-spline approximation involves the combination of a histogram and a cubic B-spline curve. To be more precise, the histogram is interpolated by a cubic B-spline curve. Figure \ref{fig:Histo} showcases this concept. 

The real probability density function is unknown, but it can be estimated using a histogram (depicted in blue). By employing a cubic B-spline curve (shown in red), a more accurate representation of the true probability density function can be achieved. This integration of a cubic B-spline curve based on a histogram effectively demonstrates the methodology behind the histogram cubic B-spline approximation.

The histogram cubic B-spline approximation relies on histograms and is influenced by the number of bins used in the histogram. Several methods exist to determine the appropriate number of bins for a histogram. One approach is to set the number of bins, denoted as $\sqrt{n}$, equal to the square root of the total number of data points. This choice aligns with the expected range of frequency fluctuations within each class and approaches $\sqrt{n}$ as the number of classes increases indefinitely. Additional approaches for determining the number of bins include Sturge's rule \cite{B7}, Scott's rule \cite{B8}, and the rule proposed by Freedman and Diaconis \cite{B9}. 

In the specific context of the demonstrated histogram cubic B-spline approximation and its validation, it is worth exploring an alternative approach known as Knuth's rule. Knuth's rule employs a Bayesian approach to determine the optimal bin width for the histogram. \cite{B6}

Each bin of the cubic B-spline curve has an arithmetic mean that matches the height of the respective bin. The following rule must be applied to the indefinite integral $F$ of the B-spline curve:
\begin{align}
F_{t_{i+1}} - F_{t_i} = h_i \left( t_{i+1}-t_{i} \right), \; i=1,\hdots,n
\end{align}
By incorporating the condition $F_{t_1} = 0$, the equations can be expressed as follows:
\begin{align}
F_{t_i} = \sum_{j=1}^{i-1} h_j \left( t_{j+1}-t_{j} \right), \; j=1,\hdots,i-1 \wedge i=1,\hdots,n+1
\end{align}
With $n$ representing the number of bins, $t_i$ denoting the left margin, $t_{i+1}$ representing the right margin of the $i$-th bin, and $h(i)$ indicating the height of the bin. \cite{MS}

The histogram cubic B-spline approximation employed in this study can also be represented using a pseudo code, as shown in Algorithm \ref{algo:pseudo}. The calculations and dimensions of the individual variables rely on the number of data points, represented as $N$, within the vector $y$ under consideration. The width of the bins, denoted as $b_w$, and the number of bins, denoted as $b_N$, are determined using a bin rule. In theory, each data point $y_i$ can be assigned a corresponding weight $w_i$. In this particular scenario, a normalized and uniform weighting scheme of $1/y_i$ is utilized. The heights of the individual bins, denoted as $H_h$, as well as the bin boundaries, denoted as $H_{(b,B)}$, are obtained from the histogram, taking into account the vector $y$, the number of bins $b_N$, and the weighting $w$. By utilizing the bin boundaries $H_{(b,B)}$, the centers of the bins, denoted as $H_c$, can also be calculated. The bin ranges, represented as $H_{b,w}$, are determined based on $H_c$. Moreover, the abscissa values, denoted as $x$, for each bin and their increments, denoted as $dx$, can be derived. The size $F$ represents the cumulative sum of $H_h \, dx$. The B-spline function $spl$ can be computed based on $x$, the size $F$, the degree $k$, and the $bc_{type}$ parameter. Its derivative is denoted as $spl_D$. Finally, the new data vector $y_{new}$ can be obtained using $spl_D$. This vector, denoted as $y_{new}$, corresponds to the B-spline curve and is associated with $x_{new}$, which includes $N$ data points. \cite{MS}

\begin{algorithm}
	\caption{Pseudo code: Histogram Cubic B-Spline Approximation} 
	\begin{algorithmic}[1]
        \State $N := len(y)$
		\State $b_w := \color{magenta} f_{bin,rule}(y) \color{black}, \; \; \; \color{blue} dim(b_w) = \#Bins+1$
        \State $b_N := \left\lfloor b_w \right\rfloor - 1, \; \; \; \color{blue} dim(b_N) = \#Bins$
        \State $w_i := 1/y_i, \; i=1,\hdots,N, \; \; \; \color{blue} dim(w) = \#y$
        \State $H_h, H_{(b,B)} := \color{magenta} histogram \color{black}(y,b_N,w), \; \; \; \color{blue} dim(H_h) = \#Bins, \color{blue} dim(b,B) = \#Bins+1$
        \State $H_c := H_{(b,B),1,...,N-1} + \color{magenta} diff \color{black}(H_{(b,B),1,...,N}) / 2, \; \; \; \color{blue} dim(H_c) = \#Bins$
        \State $H_{b,w} := \color{magenta} diff \color{black}(H_c), \; \; \; \color{blue} dim(H_{b,w}) = 1$
        \State $x :=  \left[ ( H_{c_0} - H_{b,w} / 2 ), \hdots, ( H_{c_N} + H_{b,w} / 2 ) \right], \; \; \; \color{blue} dim(x) = \#Bins+1$
        \State $dx := \color{magenta} diff \color{black}(x), \; \; \; \color{blue} \dim(dx) = \#Bins$
        \State $F := \color{magenta} cumsum \color{black} (H_h \, dx), \; \; \; \color{blue} dim(F) = \#Bins+1$
        \State \color{red} $spl := createSpline(x,F,k,bc_{type})$
        \State $spl_D := derivate(spl)$
        \State \color{black} $x_{new} :=  \left[ ( H_{c_0} - H_{b,w} / 2 ) , \hdots, ( H_{c_N} + H_{b,w} / 2 ) \right], \; \; \; \color{blue} dim(x_{new}) = \#y$
        \State $y_{new} := spl_D(_{new}), \; \; \; \color{blue} dim(y_{new}) = \#y$
	\end{algorithmic}
 \label{algo:pseudo}
\end{algorithm}

\newpage

\subsection{Validation}
The validation process aims to ensure the accuracy of the histogram cubic B-spline approximation by utilizing a relevant example within a specific application. While B-splines of higher degrees are generally applicable, cubic B-splines offer distinct advantages in this particular application due to their reduced oscillations. To validate the approximation, specific time series data related to vehicle behavior is taken into account, with a specific focus on emergency braking maneuvers. These maneuvers are examined as they represent a common safety-critical driving scenario.

During emergency braking, the longitudinal movement ($x$) of a vehicle is influenced by a range of factors including the initial speed, deceleration, driver reaction time, and the variable of time ($t$). These factors collectively contribute to the determination of the vehicle's position along its direction of motion. For a more comprehensive understanding of the underlying emergency braking, the literature \cite{pubMod} delves into further mathematical definitions and provides additional details.

\begin{figure}[H]
\centering
    \includegraphics[scale=0.88]{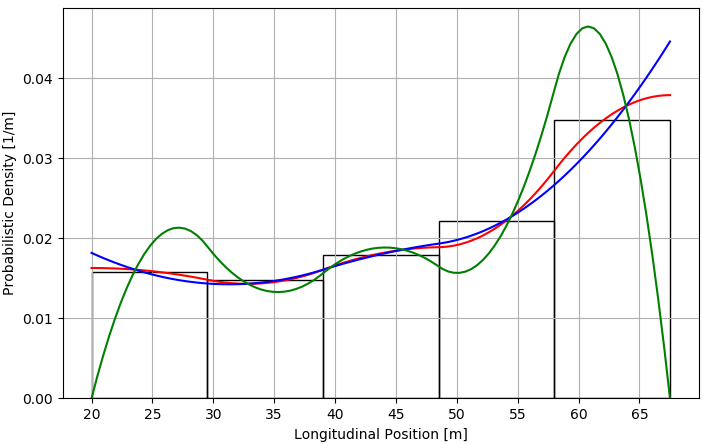}
\caption{Validation: Histogram B-Spline Approximation}
\label{fig:Graph}
\end{figure}

During the validation process, the movement of the vehicle is represented by a univariate time series, which is denoted as $x(t)$. To ensure a comprehensive validation, a total of 1000 such time series are taken into account. The validation procedure involves generating three distinct histogram cubic B-spline approximations, utilizing the provided time series data and following the algorithm described in Algorithm \ref{algo:pseudo}. These approximations are visually depicted in Figure \ref{fig:Graph}. 

It is worth noting that if a probability density function were available as a ground truth, the cubic B-spline curves could be compared to both the ground truth and each other using a statistical measure such as Kullback-Leibler divergence \cite{KL}. However, in the absence of such ground truth information, the evaluation of the approximation primarily relies on its performance and consistency within the specific application context.
 
The probability density values are plotted against the longitudinal movement (in meters) in the respective curves. The curves vary based on the type of boundary condition, denoted as $bc_{type}$ \cite{Py}.

The $bc_{type}$ \textit{clamped} (green curve) imposes zero probability density values at the ends of the histogram or respective bins. Specifically, the first derivative at the ends of the curve is zero. However, this approach does not adequately capture the characteristics of the underlying vehicle movements. In this context, each vehicle movement considered begins at $x_0 = 0$ meters and ends at $x_{end} > 65$ meters. Therefore, the corresponding probability density values should be provided by the cubic B-spline curve. That is not the case for $bc_{type}$ \textit{clamped}.

In addition to the \textit{clamped} boundary condition, the $bc_{type}$ \textit{natural} or \textit{not-a-knot} can also be utilized. The cubic B-spline curves based on the \textit{natural} condition (red curve) and the \textit{not-a-knot} condition (blue curve) exhibit similarities. For cubic B-spline curves based on the \textit{natural} condition, the second derivative at the ends of the curve is zero. For cubic B-spline curves based on the \textit{not-a-knot} condition, the first and second segments at the curve ends are the same polynomial. It is worth noting that $bc_{type}$ \textit{not-a-knot} serves as a suitable default when there is no specific information available about the boundary conditions. Due to the differing boundary conditions, there are varying numbers of turning points and curvatures for both variants. In the underlying data, the cubic B-spline curve based on the \textit{natural} condition includes three turning points, while the cubic B-spline curve based on the \textit{not-a-knot} condition only includes two turning points. As a result, the \textit{natural} variant exhibits more pronounced oscillations compared to the \textit{not-a-knot} variant. The choice between these two variants depends on the specific application and use case.

\section{Conclusion}
Random processes are frequently observed in the real world, and the underlying probability distribution of such processes is often unknown. To estimate these probability distributions, probability density functions (PDFs) can be employed. There are various methods to determine PDFs, and this publication describes a specific approach known as the histogram cubic B-spline approximation. This method combines histograms and cubic B-splines to derive PDFs. The publication also explores the impact of different boundary condition types on the resulting PDFs. The method is validated using vehicle-related time series data, specifically focusing on a safety-critical driving scenario, to demonstrate its effectiveness.

\end{document}